\documentclass[aps,prl,twocolumn,showpacs,superscriptaddress,floatfix,10pt]{revtex4-1}
\usepackage{framed}
\usepackage{graphicx}
\usepackage{amsmath}
\usepackage{epsfig}
\usepackage{helvet}
\usepackage{amssymb}
\usepackage{amsthm}
\usepackage{color}
\usepackage{enumerate}
\usepackage[scriptsize]{subfigure}
\usepackage{times}
\usepackage{dsfont}
\usepackage{bbm}
\usepackage{mathrsfs}

\def\bra#1{\mathinner{\left\langle{#1}\right|}}
\def\ket#1{\mathinner{\left|{#1}\right\rangle}}
\def\braket#1{\mathinner{\left\langle{#1}\right\rangle}}

\def\bravert{\egroup\,\vrule\,\bgroup}

\renewcommand{\H}{\mathcal{H}}

\newcommand{\si}{\sigma}
\newcommand{\ep}{\epsilon}
\newcommand{\dg}{\dagger}

\newcommand{\F}{\mathcal{F}}
\newcommand{\mm}{\mathbf{m}}
\newcommand{\pp}{\mathbf{p}}
\newcommand{\kk}{\mathbf{k}}

\newcommand{\qq}{\mathbf{q}}
\newcommand{\A}{\mathcal{A}}
\newcommand{\CC}{\mathbb{C}}

\newcommand{\id}{\mathds{1}}
\def\beq{\begin{equation}}
\def\eeq{\end{equation}}
\def\barray{\begin{eqnarray}}
\def\earray{\end{eqnarray}}

\begin{document}

\title{Many-body Lattice Wavefunctions From Conformal Blocks}

\author{Sebasti{\' a}n Montes}
\email{s.montes@csic.es}
\affiliation{Instituto de F\'{i}sica Te\'orica (IFT), UAM-CSIC, Madrid, Spain}

\author{Javier Rodr\'{\i}guez-Laguna}
\affiliation{Dept. of Fundamental Physics, Universidad Nacional de
  Educaci\'on a Distancia (UNED), Madrid, Spain}

\author{Hong-Hao Tu}
\affiliation{Max-Planck-Institut f\"ur Quantenoptik, D-85748 Garching, Germany}

\author{Germ\'an Sierra}
\affiliation{Instituto de F\'{i}sica Te\'orica (IFT), UAM-CSIC, Madrid, Spain}

\date{\today}

\begin{abstract} 
We introduce a general framework to construct many-body lattice wavefunctions starting from the conformal blocks (CBs) of rational conformal field theories (RCFTs). We discuss the different ways of encoding the physical degrees of freedom of the lattice system using both the internal symmetries of the theory and the fusion channels of the CBs. We illustrate this construction both by revisiting the known Haldane-Shastry model and by providing a novel implementation for the Ising RCFT. In the latter case, we find a connection to the Ising transverse field (ITF) spin chain via the Kramers-Wannier duality and the Temperley-Lieb-Jones algebra. We also find evidence that the ground state of the finite-size critical ITF Hamiltonian corresponds exactly to the wavefunction obtained from CBs of spin fields.
\end{abstract}



\maketitle

\section{Introduction}

Developing suitable many-body wavefunctions has provided remarkable insights into the physics of collective phenomena. Given the complexity of large quantum systems, this is a rather ardous task that demands the use of an extense theoretical arsenal. In the context of strongly-correlated systems, a tool that has yielded several fruitful results in this direction has been Conformal Field Theory (CFT) \cite{Tsvelik, QuantumGroupsCFT, diFrancesco, Gogolin, Mussardo} .



Perhaps the most notorious application may be in the realm of Fractional Quantum Hall (FQH) physics. The Laughlin wavefunction \cite{Laughlin}, used to describe an electron gas when the filling factor of the lowest Landau level is $\nu=1/n$, can be derived from a correlator of (chiral) vertex operators of a massless free boson CFT. One of the main predictions from this ansatz is the existence of Abelian fractional statistics for the emergent quasiparticles. This inspired Moore and Read to develop a general framework to describe more exotic filling fractions \cite{MooreRead91}. In particular, they proposed a Pfaffian state with topological degeneracy for $\nu=5/2$. For this state, the low-energy quasiparticles exhibit non-Abelian fractional statistics.

Progress has also been made for lattice systems. Inspired by the so-called Matrix Product States (MPS), it has been proposed that variational wavefunctions can be constructed from correlators in a CFT \cite{iMPS}. In this case, (chiral) primary operators $\phi_s(z_i)$ replace the usual finite-dimensional matrices $A_i(s)$ of the original ansatz. The most important examples have made use of Wess-Zumino-Witten (WZW) CFTs, where the internal $SU(2)_{k}$ symmetry can be exploited to encode the physical (spin) degrees of freedom. This framework has been successfully applied to systems such as the Haldane-Shastry (HS) spin chain in 1D \cite{iMPS, iMPS2} and the Kalmeyer-Laughlin (KL) model in 2D \cite{KL}. This MPS construction has also inspired sensible numerical truncation schemes for continuous FQH states \cite{Zaletel,FQH-MPS}.

One may wonder if a CFT with no internal symmetries (as, for instance, the minimal Rational Conformal Field Theories (RCFTs) originally introduced in \cite{BPZ}) can be used to generate useful lattice  wavefunctions.  Here, the only available degrees of freedom are the labels of the different fusion channels that characterize the correlators of chiral primary fields, usually known as Conformal Blocks (CBs).

The aim of this paper is to describe a general way of constructing many-body lattice wavefunctions using CBs. This can be done in a generic RCFT, regardless of the existence of an internal symmetry. After a brief introduction of the general formalism, we will illustrate these ideas revisiting the HS model and postulating a novel scheme for the well-known (chiral) Ising CFT. In the latter case, we will use the internal states of the fusion channels to encode the spin degrees of freedom. Some connections to Temperley-Lieb-Jones (TLJ) algebras \cite{QuantumGroupsCFT} will be discussed.

It should be noted that the Hilbert space that we use has already appeared in the study of Restricted Solid on Solid (RSOS) or face models \cite{QuantumGroupsCFT,IRF-DMRG}, topological quantum field theory (TQFT) \cite{NayakAnyons} and anyonic chains such as the so-called Golden Chain \cite{goldenChain,AnyonTN}. In this manuscript, our main focus will be on the analytic wavefunction amplitudes that can be obtained directly from CFT.


\section{Chiral vertex operators and Conformal Blocks}

Consider a chiral algebra $\A$ (the simplest one being the enveloping algebra of Virasoro \cite{diFrancesco}). This algebra will contain purely holomorphic fields. We can also define an analogous purely antiholomorphic algebra $\bar\A$ that commutes with $\A$. Within this formalism, the Hilbert space of a RCFT can be written as finite direct sum \cite{MooreSeiberg}
\begin{equation}
\H = \bigoplus_{i=0}^N \H_i\otimes \bar\H_i,
\end{equation}
where $\H_i$ (resp. $\bar\H_i$) is an irreducible highest-weight representation of $\A$ (resp. $\bar\A$). By definition, the representation $\H_0$ contains the identity operator and, therefore, the stress energy tensor $T(z)$ and all the operators of $\A$.

The representation $\H_i$ is infinite dimensional, but it can be split into the direct sum of finite-dimensional subspaces with a fixed value of the Virasoro operator $L_0$. Let us call $\H_i^{(0)}$ the subspace with the lowest value of $L_0$, denoted $h_i$. For minimal RCFTs, $\H_i^{(0)}$ will be a one-dimensional subspace generated by the highest-weight vector $L_0\ket{\phi_i} = h_i\ket{\phi_i}$. In the case of $SU(2)_k$ WZW models, there are $k+1$ primary fields, labeled $j=0,\frac{1}{2},\cdots,\frac{k}{2}$, such that $\H^{(0)}_j = \CC^{2j+1}$ \cite{diFrancesco}.

Different representations can be related by their fusion properties. This is summarized by the \emph{fusion coefficients} $N^i_{jk}$, which count the multiplicity of $\phi_i$ in the operator product expansion (OPE) of $\phi_j, \phi_k$, so that $\phi_j\times\phi_k=\sum_i N^i_{jk}\phi_i$. In order to simplify our discussion, we will always assume $N^i_{jk}=0,1$.

\begin{figure}[ht]
  \centering

\includegraphics[width=0.50\linewidth]{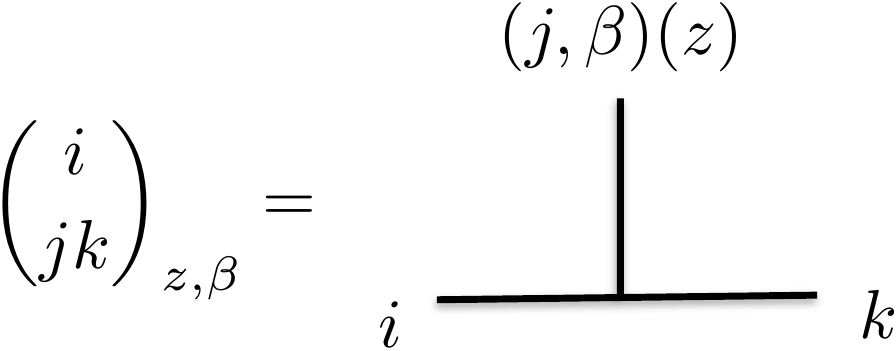}

    \caption{Graphical representation of a vertex operator. As defined in the text, $\beta\in\H^{(0)}_j$.}
    \label{vertex}
\end{figure}

We can now define \emph{chiral vertex operators} (CVO). First, consider three representations $i,j,k$ such that the fusion coefficient $N_{jk}^i$ does not vanish. A CVO is given by a linear map \cite{MooreSeiberg}
\begin{equation}
\binom{i}{jk}_{z,\beta} :\H_k^{(0)} \rightarrow \H^{(0)}_i,
\end{equation}
where $\beta\in\H^{(0)}_j$ and $z$ is a complex parameter. This can be pictured as a vertex operator with two incoming  particles with labels $j,k$ and an outgoing particle with label $i$ (Fig. \eqref{vertex}). We can also define it by the relation
\begin{equation}
\bra{\alpha}\binom{i}{jk}_{z,\beta}\ket{\gamma} = t(\alpha\otimes\beta\otimes\gamma)z^{-h_j-h_k+h_i},
\end{equation}
where $\alpha\in\H^{*(0)}_i$, $\gamma\in\H^{(0)}_k$, and $t: \H^{*(0)}_i\otimes \H_j^{(0)}\otimes \H_k^{(0)} \rightarrow \CC$ is an invariant tensor ($\H^{*(0)}_i$ is the dual of $\H^{(0)}_i$).

%

A \emph{conformal block} (CB) in a RCFT is a chiral correlator that encodes an allowed fusion channel for a given set of primary fields. If we start with $N$ primaries $\{j_n\}$, a CB can be written as \cite{QuantumGroupsCFT, MooreSeiberg}
\begin{align}
\F_\kk&(\beta_1,\cdots,\beta_N; z_1,\cdots,z_N) \\
&=\bra{0}\binom{0}{j_1 k_1}_{z_1,\beta_1}\binom{k_1}{j_2 k_2}_{z_2,\beta_2}\cdots \binom{k_{N-1}}{j_N 0}_{z_N,\beta_N}\ket{0},\nonumber
\end{align}
where  $\kk=(k_1,\cdots,k_{N-1})$ labels the internal channels. Note that $k_1=\bar j_1$ (the conjugate field of $j_1$) and $k_{N-1}=j_N$. The number of conformal blocks of this type depends on the possible allowed fusion channels of the $j_n$ fields. We will often summarize the notation and write the CB as
\begin{equation}
\F_\kk(\beta_1,\cdots,\beta_N;z_1,\cdots,z_N) = \braket{\prod_{n=1}^N \phi_{j_n}^{(\beta_n)}(z_n)}_\kk.
\end{equation}
%


\section{Lattice wavefunctions from conformal blocks}

\begin{figure}[h]
  \centering

\includegraphics[width=0.50\linewidth]{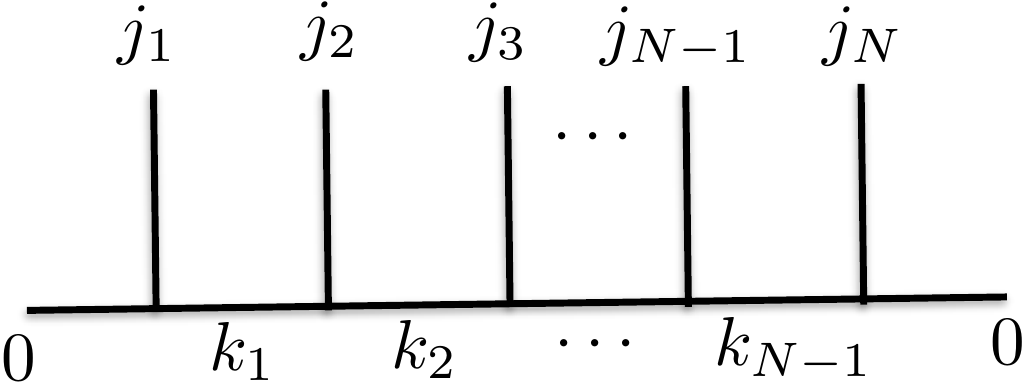}

    \caption{Graph representing a conformal block in the multiperipheral basis.}
    \label{CBgraph}
\end{figure}

For concreteness, consider a self-conjugate field $\phi$, i.e., it satisfies $\phi\times\phi = \id + \cdots$, with $\id$ being the identity field (equivalently, $N^{0}_{\phi\phi}=1$).  If we define the incidence matrix $\left(\Lambda^{(\phi)}\right)^i_j=N^{i}_{\phi \,j}$ , the number of different CB obtained from $n$ $\phi$ fields is \cite{QuantumGroupsCFT}
\begin{equation}
d_{\phi, n} = \left(\left[\Lambda^{(\phi)}\right]^n\right)^0_0,
\end{equation}
where $0$ stands for the $\id$ component. The canonical basis for these CBs uses $\kk$ for the labeling and is called the \emph{multiperipheral basis} \cite{MooreSeibergNotes}. It is closely related to the string Hilbert spaces used in RSOS or face models \cite{QuantumGroupsCFT,IRF-DMRG} (see Fig. \eqref{CBgraph}).

We will distinguish two cases:

\begin{enumerate}

\item \emph{Abelian:} if $d_{\phi, n}=1$, there is only one CB. It defines a map
\begin{equation}
\F: \H^{(0)}_\phi\otimes\cdots\otimes\H^{(0)}_\phi \rightarrow \CC,
\end{equation}
that allows us to postulate, for a fixed set of coordinates $\{z_i\}$, the wavefunction
\begin{equation}
\ket{\psi} = \sum_{\{s_i\}}\F(s_1,\cdots,s_n)\ket{s_1,\cdots,s_n},
\end{equation}
where $\{\ket{s}:s=1,\cdots,\dim(\H^{(0)}_\phi)\}$ is an orthonormal basis for $\H^{(0)}_\phi$.

\item \emph{Non-Abelian:} if $d_{\phi, n} > 1$, the different CBs must be labelled by the internal fusion channels. This defines a family of maps
\begin{equation}
\F_\kk: \H^{(0)}_\phi\otimes\cdots\otimes\H^{(0)}_\phi \rightarrow \CC,
\label{CBbasis}
\end{equation}
that suggests the wavefunction
\begin{equation}
\ket{\psi} =\sum_{\{s_i\},\kk}\F_\kk(s_1,\cdots,s_n)\ket{\kk}\otimes\ket{s_1,\cdots,s_n},
\label{NonAbelianPsi}
\end{equation}
where $\ket{\psi}\in W_{\phi,n}\otimes\left(\H^{(0)}_\phi\right)^{\otimes n}$, and $W_{\phi,n}$ contains all the auxiliary degrees of freedom (note that $\dim(W_{\phi,n})=d_{\phi,n}$).

Consider the case when $\dim(\H^{(0)}_\phi)=1$. Wavefunction \eqref{NonAbelianPsi} can represent a many-body lattice system if the auxiliary Hilbert space $W_{\phi,n}$ can account for all the local physical degrees of freedom. It should be noted that there are different sensible bases for this auxiliary space. These different representations are usually related to the order in which we fuse the primary fields. This implies that there are several lattice wavefunctions that can be obtained from the same CBs in the non-Abelian case. The most natural option is the multiperipheral basis. However, there are other bases that can be chosen for physical reasons. We will exemplify this later in this paper.

One consistency condition we need to check is that the normalization of the state has no monodromy issues with respect to the auxiliary coordinates $\{z_i\}$. This can be stated in terms of the full correlator
\begin{equation}
\braket{\psi | \psi} = \braket{\phi(z_1,\overline{z_1})\cdots\phi(z_n,\overline{z_n})} = \sum_\kk d_\kk \overline{\F_{\kk}}\F_{\kk},
\label{NormCondition}
\end{equation}
where $d_\kk$ are constants independent of $\{z_i\}$ that can be computed from the fusion matrix \cite{MooreSeibergNotes}. In the multiperipheral basis, this implies the necessary condition $\braket{\kk|\kk'}=d_\kk \delta_{\kk,\kk'}$. 
\end{enumerate}

We will illustrate this general construction with two examples. First, the Haldane-Shastry spin chain from the point of view of the $SU(2)_1$ WZW model. This case has already been studied extensively \cite{iMPS, iMPS2}. We will provide a brief summary for the sake of clarity and analogy. Then, we will introduce the use of CBs in the context of the Ising model. Being non-Abelian and lacking internal symmetry, it provides a good testing ground for encoding physical degrees of freedom using only the internal fusion channels.


\section{Abelian case study: The Haldane-Shastry chain}

The chiral algebra of the $SU(2)_k$ WZW model is the Kac-Moody algebra defined by the conserved chiral currents $J^a(z)$. It contains a representation of the Virasoro algebra that can be obtained from the stress-energy tensor $T(z)$ via the relation
\begin{equation}
T(z) = \frac{1}{2(k+2)} \sum_a (J^a J^a)(z)
\end{equation}
using the Sugawara construction \cite{diFrancesco}. For this family of models, the conformal symmetry is enriched, allowing for heighest-weight representations with more structure.

\begin{figure}[ht]
  \centering

\includegraphics[width=0.80\linewidth]{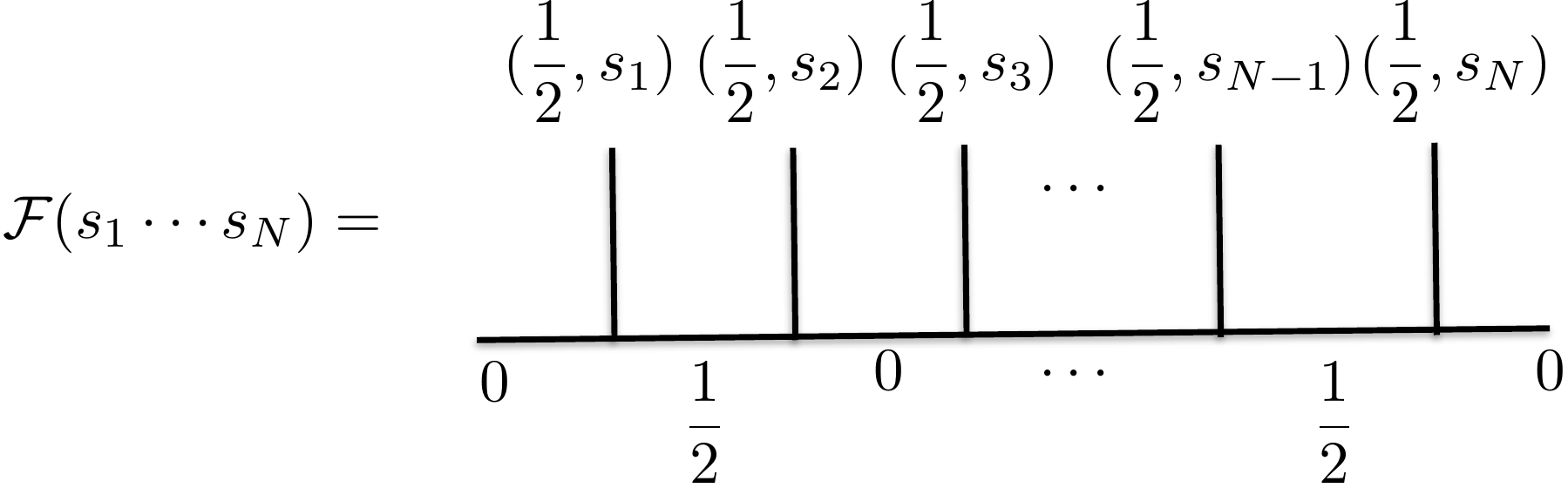}

    \caption{Graphical representation of the CB for $N$ $\phi_{1/2}$ fields.}
    \label{HS-CB}
\end{figure}

Consider $SU(2)_1$. This theory has two primary operators, $\phi_0$ and $\phi_{1/2}$, satisfying the fusion rules
\begin{equation}
\phi_{1/2}\times \phi_{1/2} = \phi_0, \quad \phi_{1/2}\times \phi_0=\phi_{1/2}, \quad \phi_0\times\phi_0=\phi_0.
\end{equation}
The primary field $\phi_{1/2}$ has a spin-$\frac{1}{2}$ representation, so that $\H^{(0)}_{1/2}=\CC^2$. From the fusion rules, we obtain the set of CVOs
\begin{equation}
\phi_{1/2,\text{odd}}^{(s)}(z)=\binom{0}{\frac{1}{2}\frac{1}{2}}_{z,s}, \quad \phi_{1/2,\text{even}}^{(s)}(z)=\binom{\frac{1}{2}}{\frac{1}{2}0}_{z,s},
\end{equation}
where we use the third component of the spin $s=\pm 1$ to label the internal degree of freedom. The CB of $N$ $\phi_{1/2}$ fields will alternate both even and odd CVOs (see Fig. \eqref{HS-CB}). This implies that there will only be a single internal fusion channel. Note also that, given the fusion rules, we will only obtain non-trivial results if $N$ is an even number.

The value of the CB can be easily computed \cite{iMPS, iMPS2}
\begin{equation}
\F(s_1,\cdots,s_N) = \rho_{1/2}\prod_{i>j}\left(z_i - z_j\right)^{s_i s_j/2},\quad \sum_i s_i = 0,
\label{CB SU(2)_1}
\end{equation}
where $\rho_{1/2} = \exp\left(\frac{i\pi}{2}\sum_{i\text{ odd}}(s_i-1)\right)$ is a Marshall sign factor. This wavefunction corresponds to the HS state for the choice $z_n = \exp(\frac{i2\pi}{N}n)$
\begin{equation}
\psi_{s_1\cdots s_N} \propto \rho_{1/2} \prod_{n>m}\left[\sin\left(\frac{\pi(n-m)}{N}\right)\right]^{s_n s_m/2}.
\label{HSwavefunction}
\end{equation}

The constraints imposed by the current algebra can be exploited to obtain parent Hamiltonians. The general construction relies heavily on the fusion properties of the different representations of the primary fields \cite{iMPS2}. For wavefunction \eqref{HSwavefunction}, the parent Hamiltonian can be related to the HS Hamiltonian \cite{iMPS}
\begin{equation}
H_{HS} = - \sum_{i\neq j}\frac{z_i z_j}{(z_i-z_j)^2}\left(\vec{\sigma}_i \cdot \vec{\sigma}_j - 1\right).
\end{equation}
%


\section{Non-Abelian case study: Ising model}

The (chiral) Ising CFT is a minimal RCFT, so the corresponding chiral algebra will be the (enveloping) Virasoro algebra \cite{Mussardo}. It consists of three primary fields, $\id$, $\psi$ (Majorana) and $\sigma$ (spin) with conformal weights $0$, $1/2$ and $1/16$, respectively. They have the (non-trivial) fusion rules \cite{diFrancesco}
\begin{equation}
\sigma\times\sigma = \id + \psi, \qquad \psi\times\psi = \id, \qquad \sigma\times\psi = \sigma,
\end{equation}
which imply $N^\id_{\sigma\sigma} = N^\psi_{\sigma\sigma} = N^\sigma_{\sigma\id} = N^\sigma_{\sigma\psi}=1$.

The use of CBs obtained from the Ising CFT has been popular in the context of FQH physics ever since the seminal paper by Moore and Read \cite{MooreRead91, NayakWilczek}. They usually represent a ground-state given by a Pfaffian obtained from a condensate of Majorana fields. Quasiholes are represented by $\sigma$ fields that create topological degeneracies. Up to some confining factors, these wavefunctions can be written as
\begin{equation}
\Psi_\pp \propto \braket{\prod_{i=1}^N \psi(z_i)\prod_{j=1}^{2m}\sigma(w_j)}_\pp,
\end{equation}
where $\pp$ stands for the different fusion channels of the $\sigma$ fields. In this context, the complex coordinates stand for the physical positions of the different particles in the condensate. The internal fusion channels correspond to globally distinct topological sectors. In this manuscript, we will interpret the available degrees of freedom in a different way. We will encode lattice spin degrees of freedom using the different fusion channels of the $\sigma$ fields. The complex coordinates can then be related to variational parameters.

The general formulas for the CBs obtained from the Ising primary fields have been calculated in \cite{IsingCB}. We will be particularly interested in the CBs containing only $2N$ spin field operators
\begin{equation}
\F^{(2N)}_\pp (z_1,\cdots,z_{2N}) = \braket{\si(z_1)\cdots\si(z_{2N})}_\pp,
\end{equation}
where $\pp$ labels the corresponding internal fusion channels. Note that, being a minimal RCFT, $\dim(\H^{(0)}_\sigma) = 1$.

\begin{figure}[h]
  \centering

\includegraphics[width=0.90\linewidth]{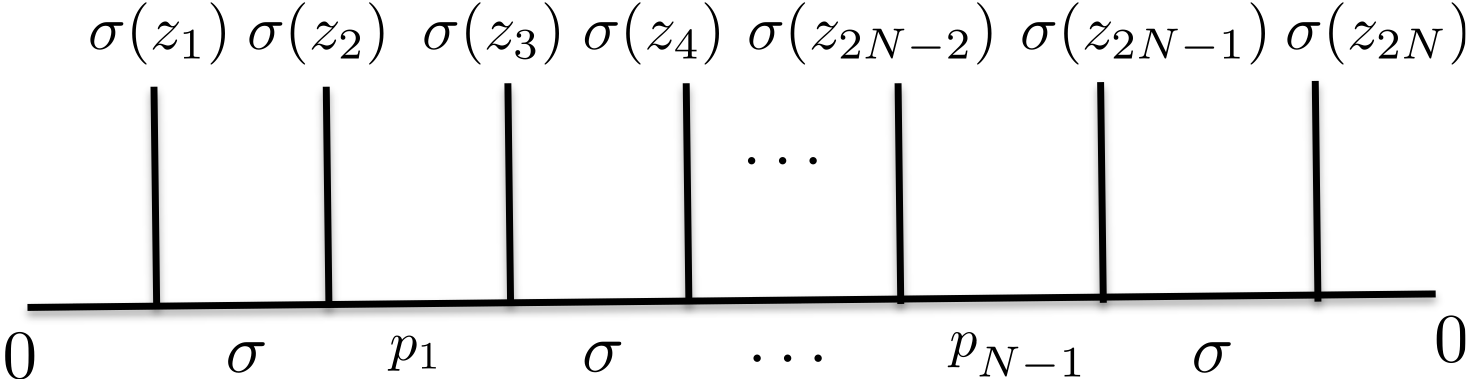}

    \caption{Graphical representation of $\F_\pp^{(2N)}$ in the multiperipheral basis. }
    \label{SigmaCB}
\end{figure}

It is easy to show that $2N$ $\sigma$ fields generate $2^{N-1}$ different CBs. One way of understanding this is by considering each $\sigma$ field as half of a binary degree of freedom. This is consistent with the fact that the so-called quantum dimension of $\sigma$ is $d_\sigma=\sqrt{2}$ \cite{QuantumGroupsCFT,  NayakAnyons}. On top of this, there is a (fermionic) parity constraint imposed by the Majorana fields that reduces in half the dimension of the Hilbert space.

Given the fusion rules, some internal channels will be fixed to $\sigma$ while others can be either $\id$ or $\psi$ (see Fig.\eqref{SigmaCB}). We will omit the $\sigma$'s in this multiperipheral representation and write $\pp=(p_1,\cdots,p_{N-1})$, where $p_i=0$ (resp. $1$)  corresponds to an identity operator $\id$ (resp. a fermion $\psi$).  (Note that $\pp$ can take $2^{N-1}$ different values, as expected.)

\begin{figure}[ht]
  \centering

\includegraphics[width=0.90\linewidth]{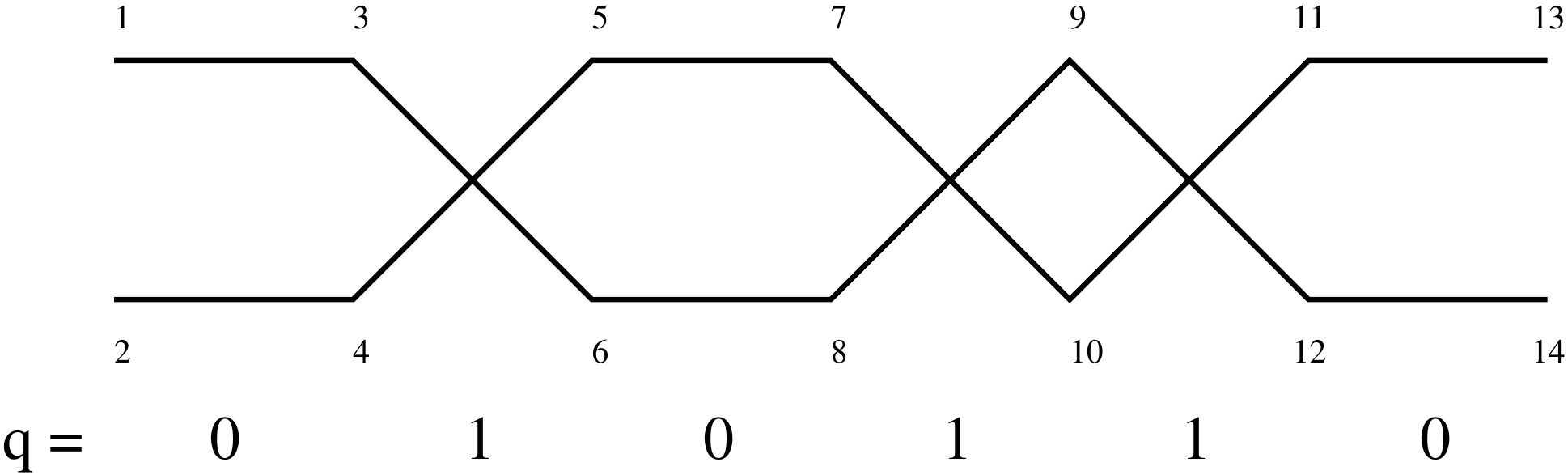}

    \caption{Graph representing one of the possible pairs of macrogroups for 14 $\si$ fields, corresponding to $\ell_\qq = (1, 3, 6, 8, 9, 12,14)$ and $\ell'_\qq = (2, 4, 5, 7, 10, 11, 13)$. }
    \label{MGgraph}
\end{figure}

Before stating the formulae for the CBs, we introduce some extra notation. First, we will need certain bipartitions of the $\si$ field coordinates that associate the points of each reference pairs to different groups. We call these \emph{macrogroups} $\ell_\qq, \ell'_\qq$ and they are generated from an integer $\qq=0,\cdots, 2^{N-1}-1$ according to the algorithm \cite{NayakWilczek, IsingCB}
\begin{align}
&\ell_\qq(1) = 1, \nonumber \\
&\ell_\qq(k+1) - \ell_\qq(k) = 2 \quad \text{if} \; q_k = 0, \\
&\ell_\qq(k+1) - \ell_\qq(k) = 1 \quad \text{if } q_k = 1 \text{ and } \ell_k \text{ is even},\nonumber \\
&\ell_\qq(k+1) - \ell_\qq(k) = 3 \quad \text{if } q_k = 1 \text{ and } \ell_k \text{ is odd},\nonumber
\end{align}
where $q_k$ are the binary digits of $\qq = (q_1, q_2, \dots, q_{N-1})$. The macrogroup $\ell'_\qq = ({\ell'_\qq(1)}, \cdots, {\ell'_\qq(N)})$ associated to $\qq$  satisfies the same recursion relations with the initial condition  $\ell'_\qq(1) = 2$. We can get an idea of these type of bipartitions from a graphical representation (Fig.\ref{MGgraph}).

Note that we can also rewrite the macrogroup numbers in an economical way
\begin{equation}
\ell_\qq(k) = 2k -\frac{1}{2}(1+s_k), \qquad \ell'_\qq(k) = 2k -\frac{1}{2}(1-s_k),
\label{Spin macrogroup}
\end{equation}
where
\begin{equation}
s_k = \prod_{i=1}^{k-1}\left(1-2q_i\right),
\end{equation}
and $s_1=1$ by definition.

Using this notation, we can define
\begin{equation}
z_{\ell_\qq}=\prod_{k>m}z_{\ell_\qq(k),\ell_\qq(m)}.
\end{equation}
where $z_{ab} = z_a - z_b$. We will also need the sign given by
\begin{equation}
\ep_{\pp\qq} = (-1)^{\sum_k p_k q_k},
\label{eps}
\end{equation}
using the binary expansion of both $\pp$ and $\qq$.

The expression for the CB can be written as \cite{IsingCB}
\begin{equation}
\F^{(2N)}_\pp = \frac{1}{2^{\frac{N-1}{2}}}\prod_{a>b}^{2N}z_{ab}^{-1/8}\left(\sum_{\qq = 0}^{2^{N-1}-1}\ep_{\pp\qq}\sqrt{z_{\ell_\qq}z_{\ell'_\qq}}\right)^{1/2}.
\label{CBsigmas}
\end{equation}
Note that the sum inside the square root is the only part that depends on $\pp$. 

We will focus now on a one-dimensional configuration. The coordinates are chosen to be on the unit circle
\begin{equation}
z_n = \exp\left[i \frac{2\pi}{2N}\left(n+(-1)^n\delta\right)\right]=\exp\left[i\theta_n(\delta)\right],
\label{coord}
\end{equation}
where $-\frac{1}{2}<\delta<\frac{1}{2}$ is a variational parameter.

>From \eqref{CBsigmas}, we note that we can ignore all the prefactors that do not depend on $\pp$. Using the familiar identity
\begin{equation}
z_n -z_m = 2 i \exp\left(i\frac{\theta_n+\theta_m}{2}\right)\sin\left(\frac{\theta_n-\theta_m}{2}\right)
\label{ComplexToSine}
\end{equation}
the normalized wavefunction amplitudes can be rewritten using only positive weights (we neglect an overall phase)
\begin{equation}
\Psi_\pp(\delta) = \frac{1}{\tilde N_0}\left(\sum_{\qq = 0}^{2^{N-1}-1}\ep_{\pp\qq}\, A_\qq(\delta)\right)^{1/2},
\label{varWF}
\end{equation}
where
\begin{equation}
A_\qq = \prod_{n>m}^N\left[\sin\frac{\theta_{\ell_\qq(n)}-\theta_{\ell_\qq(m)}}{2}\sin\frac{\theta_{\ell'_\qq(n)}-\theta_{\ell'_\qq(m)}}{2}\right]^{\frac{1}{2}},
\label{AAA}
\end{equation}
and
\begin{equation}
\tilde N_0^2 = \frac{N^{N/2}}{2^{(N-1)(N-2)/2}}.
\end{equation}
Note that $A_0$ and the normalization constant $\tilde N_0$ do not depend on $\delta$ (see Appendix A). For each amplitude, there is an exponentially large number of numerical operations that need to be performed in order to compute it. This imposes some size contraints on the possible numerical implementations of this ansatz written in this form.


\section{Braiding and Kramers-Wannier duality}

In the non-Abelian case, the braiding of fields will in general mix different CBs, $\F_{\kk} = \sum_{\kk'}B_{\kk,\kk'}\F_{\kk'}$, where $B_{\kk,\kk'}$ is a representation of the braiding operation \cite{MooreSeibergNotes, QuantumGroupsCFT}.  Geometrically, we expect that cyclic permutation of primary fields located on the unit circle according to \eqref{coord} will yield useful algebraic properties.

Assume there are $M$ primary fields $\phi$. If we take $\phi(z_{M})$ and braid it with all the other fields, we obtain the same CB up to a relabeling $z_{k}\to z_{k+1}$ (identifying $M+1\equiv 1$), so that
\begin{equation}
\F_\pp(\delta) = \sum_{\pp'} U_{\pp \pp'}\F_{\pp'}(-\delta),
\label{F-KW}
\end{equation}
where $U$ is the operator obtained from the braiding.

This process can be decomposed into pair-wise permutations. Call $\omega_i$ the operator that interchanges $\phi(z_i)$ and $\phi(z_{i+1})$.  The set $\{\omega_i|i=1,\cdots,M-1\}$ will satisfy the \emph{braid group} relations
\begin{equation}
\omega_i\omega_{i+1}\omega_i = \omega_{i+1}\omega_i\omega_{i+1}, \quad \omega_i\omega_j = \omega_j\omega_i, \,|i-j|\geq 2.
\end{equation}
It can be shown that \cite{MooreSeibergNotes}
\begin{equation}
\omega_1\cdots \omega^2_{M-1}\cdots \omega_1 = \exp\left( -4i\pi\Delta_\phi\right)
\end{equation}
where $\Delta_\phi$ is the scaling dimension of $\phi$ and the sign of the phase depends on the direction of the exchange. This allows us to define the unitary operator
\begin{equation}
U = e^{2i\pi\Delta_\phi} \omega_1\cdots \omega_{M-1}=e^{-2i\pi\Delta_\phi} \omega_1^\dg\cdots \omega_{M-1}^\dg,
\end{equation}
which is independent of the convention for the pair-wise exchange.

\begin{figure}[h]
  \centering

\includegraphics[width=1.00\linewidth]{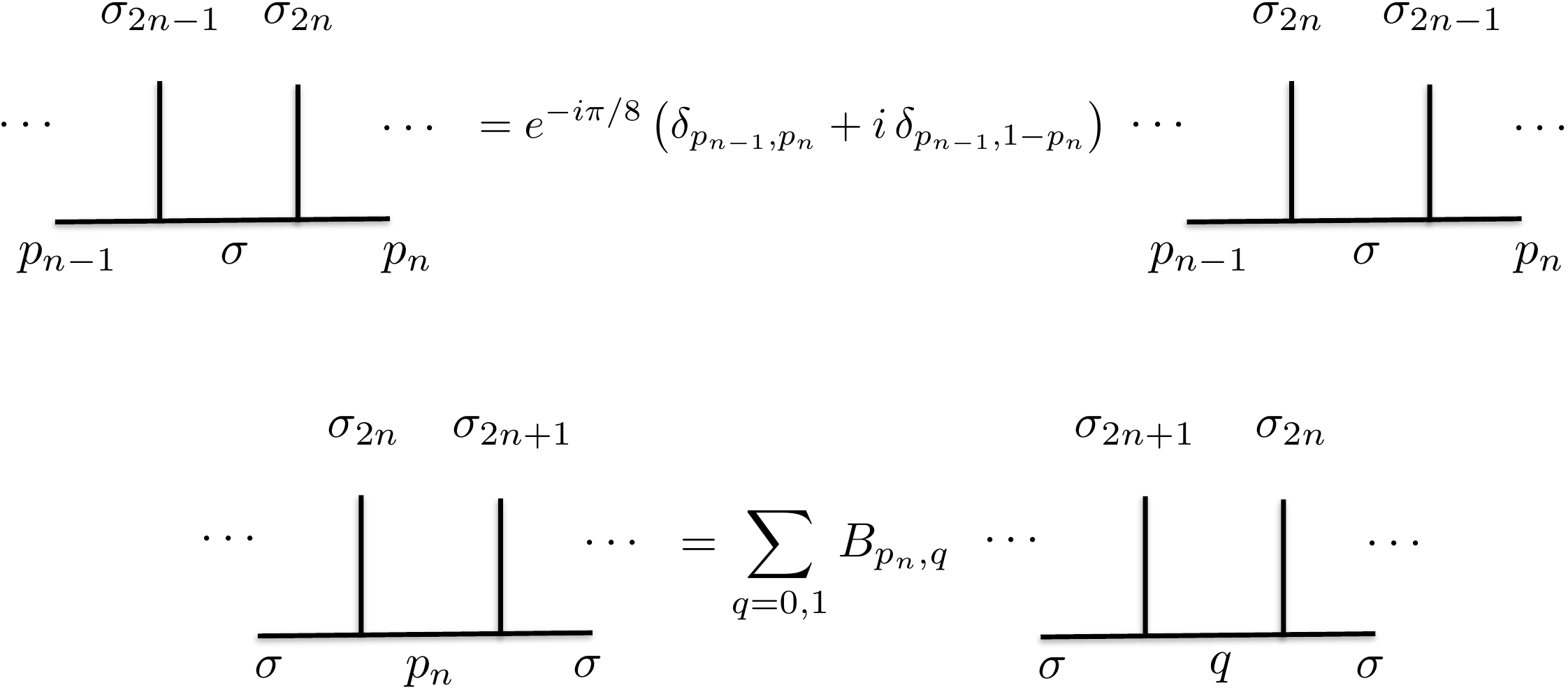}

    \caption{Braiding operators for $\F^{2N}$. We define $B=\frac{e^{i\pi/8}}{\sqrt{2}}\begin{pmatrix}
  1 & i \\
  i & 1
 \end{pmatrix}$ and use the convention $\sigma_i = \sigma(z_i)$.}
    \label{IsingBraiding}
\end{figure}

We can obtain a representation of these operators from the fusion matrix $F$ and the braid operator $R$ \cite{QuantumGroupsCFT, NayakAnyons}. For $2N$ $\sigma$ fields, they can be constructed from two local unitary operators (see Fig. \eqref{IsingBraiding}).

The action of this operation can be better understood using a different, dual basis. First, we group the coordinates in \emph{reference pairs} $(\si(z_{2k-1}), \si(z_{2k}))$. If they are fused pairwise, the different channels can be labeled using the vector $\mm=(m_1,\cdots,m_N)$, with $m_i=0$ (resp. $1$) representing an identity operator $\id$ (resp. a fermion $\psi$). In this case, there is an enforced parity coming from the preservation of fermion parity, so that the dimension of the Hilbert space stays the same.

\begin{figure}[h]
  \centering

\includegraphics[width=0.95\linewidth]{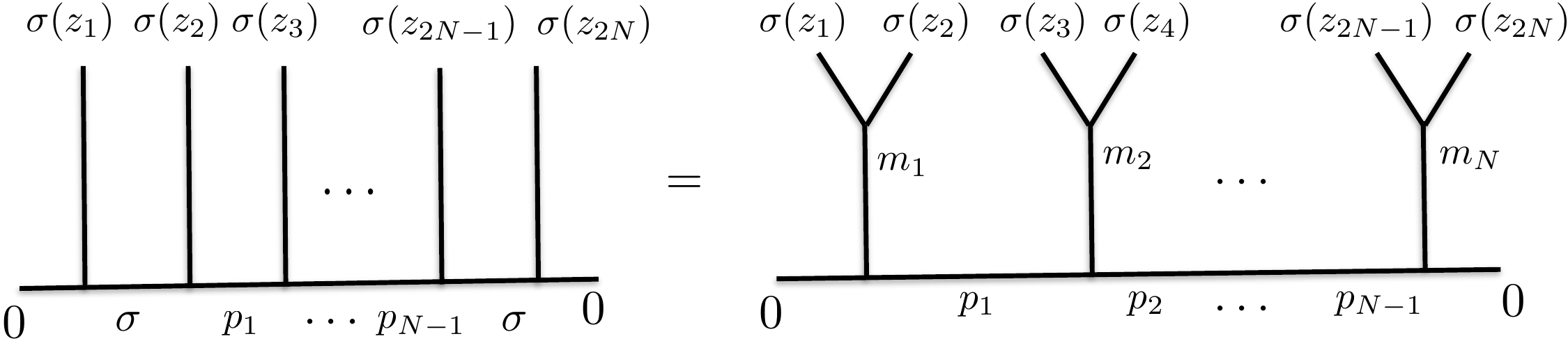}

    \caption{A conformal block using only $\si$ field operators grouped in reference pairs $(\si(z_{2k-1}), \si(z_{2k}))$. The equivalance between the two representations is obtained from the relation $m_k = p_{k-1} + p_k (\text{mod}\, 2)$.}
    \label{IsingCB}
\end{figure}

Both bases can easily be related (see Fig. \eqref{IsingCB}). Note that, in order to preserve the number of fermions at each vertex, there is the restriction $m_k = p_{k-1} + p_k (\text{mod}\, 2)$. (We define fixed auxiliary values $p_0=p_N=0$.)

In this basis, the operator $U$ corresponds to the Kramers-Wannier transformation restricted to the even parity sector ($\braket{\prod_n \sigma^z_n} = 1$)
\begin{equation}
U\sigma^z_n U^\dg = \sigma^x_n\sigma^x_{n+1}, \quad U\sigma^x_n U^\dg = \sigma^z_1\cdots\sigma^z_{n}.
\end{equation}
This duality is known to be non-invertible in the odd parity sector \cite{topDefIsing}.

Note also that iterating this process, i.e. braiding the whole last reference pair, corresponds to a one-site translation in the pair-wise fusion basis.

Relation \eqref{F-KW} implies that $\F(\delta=0)$ will be self-dual. This is also the case for the (even parity) ground state of the critical Ising Transverse Field (ITF) Hamiltonian
\begin{equation}
H = -\sum_{n=1}^N \sigma^x_n\sigma^x_{n+1} - \sum_{n=1}^N \sigma^z_n,
\label{ITFh1}
\end{equation}
as a consequence of the self-duality of the critical Ising model. As we will see numerically, this is more than a mere coincidence.


\section{The Temperley-Lieb-Jones algebra}


We can associate an integrable model to a given minimal RCFT using the properties of the CBs. This is due to the fact that the constraints imposed by the braiding and fusion of operators can be related to the Yang-Baxter algebra \cite{QuantumGroupsCFT}.

We first need some general definitions. A \emph{Temperley-Lieb-Jones} (TLJ) algebra is an unital, 
complex algebra closely related to the braid group. It is generated by operators $\{e_i | i=1,\cdots, M\}$ satisfying
\begin{align}
e_i^\dagger  = e_i, &\qquad e_i e_{i\pm 1} e_i = e_i ,\\
e_i^2 = \sqrt{\beta} e_i, &\qquad e_i e_j  = e_j e_i, \quad \text{for } |i-j|\geq 2,\nonumber
\end{align} 
where $\beta$ is a free parameter which for the Ising model takes the value  $\beta_{\text{Ising}}=2$. 
Let $\{\ket{\pp} = \ket{p_1, \dots, p_{N-1}} \}$ be the basis of the Hilbert space of the  model.
The action of the TLJ operators on  this basis is given by  \cite{QuantumGroupsCFT}
\barray 
e_{2 n -1} | {\bf p} \rangle  & = & \sqrt{2} \; \delta_{p_{n-1}, p_n}  \,|  \dots,  p_{n},  \dots  \rangle , \;  n=1, \dots, N
\label{T2}   \\
e_{2 n } | {\bf p} \rangle  & = & \frac{1}{\sqrt{2}}  \left(  \, |  \dots,  p_{n-1}, p_{n},  \dots  \rangle \right.   \nonumber \\
&  + &  \left.   |  \dots, p_{n-1}, 1-p_{n}, \dots  \rangle \right), \; n=1, \dots, N-1   \, ,    \nonumber  \\
e_{2 N} | {\bf p} \rangle  & = & \frac{1}{\sqrt{2}}  \left( | {\bf p} \rangle  + | {\bf p'} \rangle  \right) \, ,   \nonumber 
\earray  
where $\pp' = (1-p_1, \dots, 1- p_{N-1})$.  The last element,  $e_{2 N}$,  is  an extension of the TLJ algebra
to periodic systems  \cite{PTLJ}. The TLJ operators can also be expressed in the spin basis  $\{\ket{\mm}  =
\ket{m_1, \dots, m_N} \}$,  where $m_k = p_k + p_{k-1} \; ({\rm mod} \; 2)$ (recall Fig. \ref{IsingCB}), 
\barray
\sqrt{2} \, e_{2 n-1} - 1 & = & \sigma^z_n, \quad  n = 1, \dots, N ,  \label{T3} \\
\sqrt{2} \, e_{2 n} - 1 & = & \sigma^x_n \sigma^x_{n+1}, \quad  n = 1, \dots, N \, ,   \nonumber 
\earray 
with the periodicity conditions $\sigma^x_{N+1} = \sigma^x_1$. 
One can verify that  the operator  $U$, 
that implements the KW duality, satisfies 
\begin{equation}
e_{n+1} = U e_n U^\dg, \quad n=1,\cdots 2N.
\end{equation}
%
%
%
%
%
Using equations \eqref{T3}, it is clear that the critical ITF Hamiltonian \eqref{ITFh1} corresponds to the Temperley-Lieb Hamiltonian 
\begin{equation}
H_{TL}  = - \sum_{i=1}^{2 N} \left(  \sqrt{2} \,  e_i - 1 \right).
\label{HamTL}
\end{equation} 

We expect then that the many-body state constructed from the CBs of the Ising model, namely
\beq
\ket{\psi} = \sum_\pp \F_\pp \ket{\pp},
\label{T4}
\eeq
will be closely related to the spin chain Hamiltonian \eqref{HamTL}. Since this  Hamiltonian is translational invariant, one is lead to the choice  $z_n = e^{ 2 \pi i n/(2N)}$. This guarantees that the state \eqref{T4} is both
translational invariant and KW self-dual. The eigenvalue equation $H |\psi \rangle = E | \psi \rangle$ reads explicitly
\barray 
- \sum_{n=1}^N ( 2 \delta_{p_{n-1}, p_n} -1) \F_{\dots, p_n, \dots} 
- \sum_{n=1}^{N-1} \F_{\dots, 1- p_n, \dots}  \label{T5} \\
- \F_{1-p_1, \dots, 1- p_{N-1}}  = E \, \F_{\dots, p_n, \dots} 
\nonumber 
\earray 
We  shall show  below  that this equation holds numerically and that \eqref{T4}  coincides with the ground state of the critical Ising model. This result does not follow from the algebraic construction presented above, although it serves as a motivation. In the case of the Haldane-Shastry wave function,  the associated parent Hamiltonian was derived using the properties of null vectors of the Kac-Moody algebra $SU(2)_1$  \cite{iMPS,iMPS2} .
Here too we expect equation \eqref{T5} to follow from the null vectors of the Virasoro module (we leave this discussion for a later publication).


\section{Characterization of the variational wavefunctions}

Consider $\delta\to -1/2$. From the OPE, we know that the identity will be more dominant than the fermion in each reference pair. Since in this limit $z_{2k-1}\to z_{2k}$, we have that
\begin{equation}
z_{\ell_\qq}z_{\ell'_\qq} \to z_{\ell_\qq}^2=z_{\ell_0}^2.
\end{equation}
so
\begin{equation}
\Psi_\pp\left(\delta\to -\frac{1}{2}\right) \to \delta_{\pp,0},
\end{equation}
i.e., a trivial product state $\ket{\Psi(\delta\to -\frac{1}{2})}=\ket{0}$.

Now, take $\delta\to 1/2$. In this limit, $z_{2k}\to z_{2k+1}$. Given that for $\qq\neq 0$ there will be at least one difference that vanishes (for instance, if $q_1=1$, then $z_{\ell'_\qq}=0$ because it contains the factor $z_2-z_3$), we have
\begin{equation}
z_{\ell_\qq}z_{\ell'_\qq} \to \delta_{\qq,0}\,z_{\ell_0}z_{\ell'_0}.
\end{equation}
This implies that all the configurations have equal weight. In the pair-wise fusion basis, this is a ferromagnetic state projected onto the even parity sector, i.e.,
\begin{equation}
\ket{\Psi\left(\delta\to \frac{1}{2}\right)} \to \frac{1}{\sqrt{2}}\left(\ket{+}^{\otimes 
N}+\ket{-}^{\otimes N}\right),
\end{equation}
where $\sigma^x\ket{\pm}=\pm\ket{\pm}$.

These two limiting cases correspond to the trivial phases of the Ising transverse field (ITF) spin chain
\begin{equation}
H(h) = -\sum_{n=1}^N \sigma^x_n\sigma^x_{n+1} -h \sum_{n=1}^N \sigma^z_n,
\label{ITFanyH}
\end{equation}
i.e., $h\to\infty$ and $h=0$, respectively, in the even parity sector, meaning $\braket{Q}=\braket{\prod_n \sigma^z_n}=1$. Moreover, both are dual after a Kramers-Wannier transformation, as expected from \eqref{F-KW}.

Given these particular cases, it is tempting to relate the varional parameter $\delta$ to the external magnetic field $h$. In order to study this, we computed for different system sizes the value of $h$ that maximizes the overlap between $\ket{\Psi(\delta)}$ and the ground state $\ket{gs(h)}$ . (We limit our analysis to $0\leq \delta\leq \frac{1}{2}$. Negative values of $\delta$ have a similar behavior due to the KW duality.)

\begin{figure}[h]
  \centering

\includegraphics[width=0.75\linewidth]{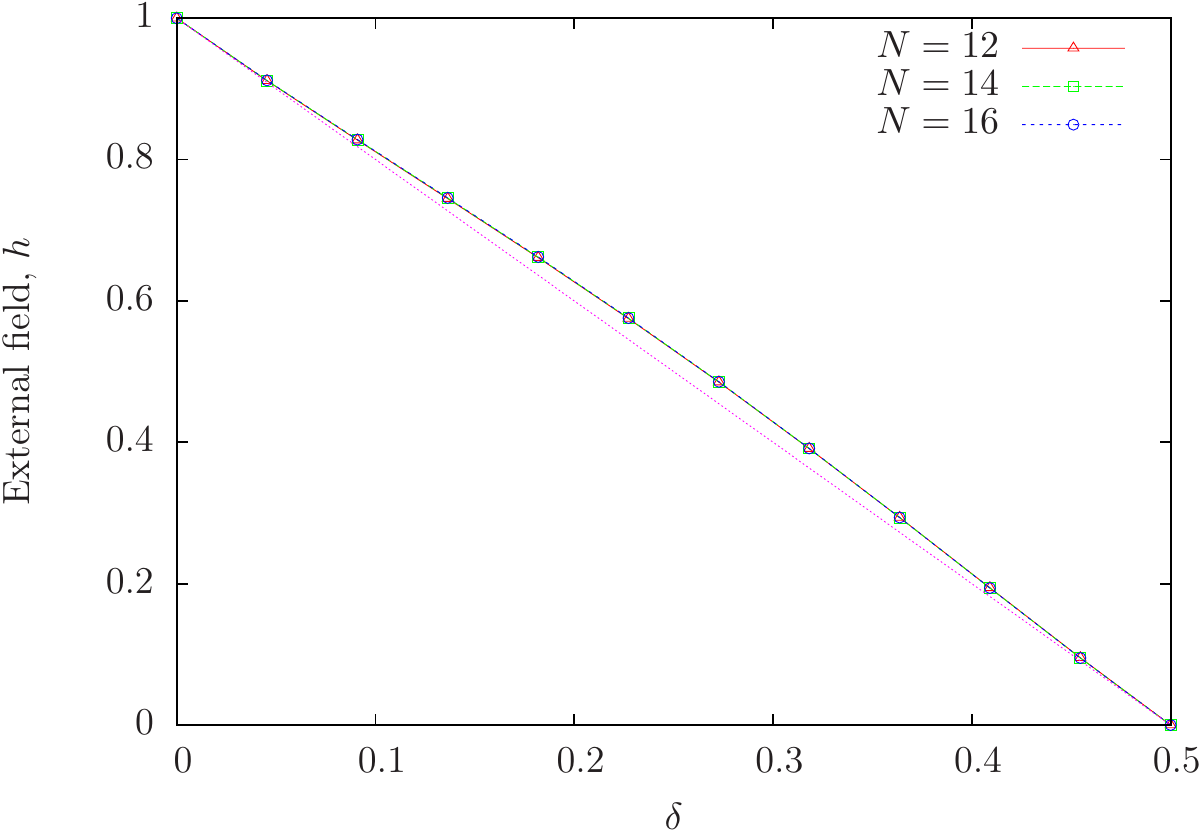}

    \caption{Relation between the variational parameter $\delta$ and the value of the external field $h$ that optimizes the overlap. (The dotted purple line is a reference straight line.) Note that the behavior is virtually independent of the system size. }
    \label{map}
\end{figure}

\begin{figure}[h]
\centering
\subfigure[]{
   \includegraphics[width=0.85\linewidth]{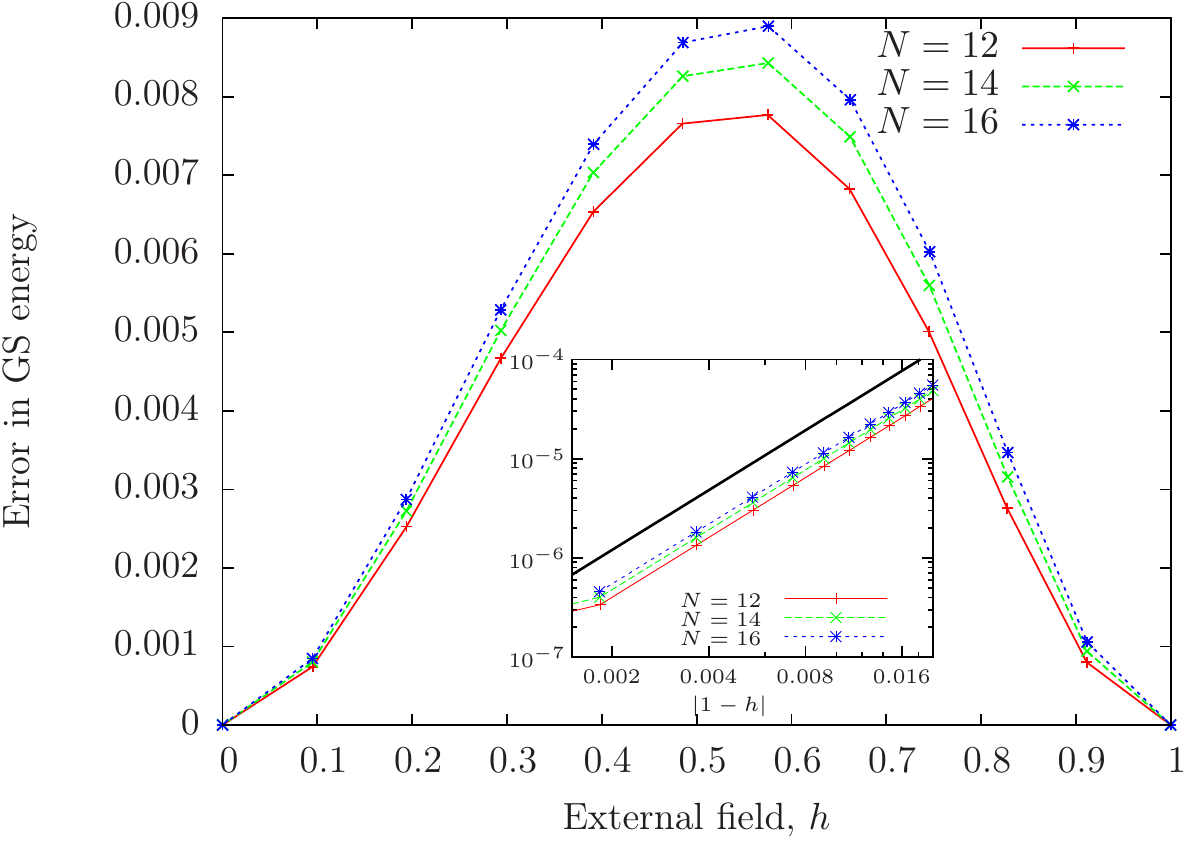}
   }
\subfigure[]{
   \includegraphics[width=0.85\linewidth]{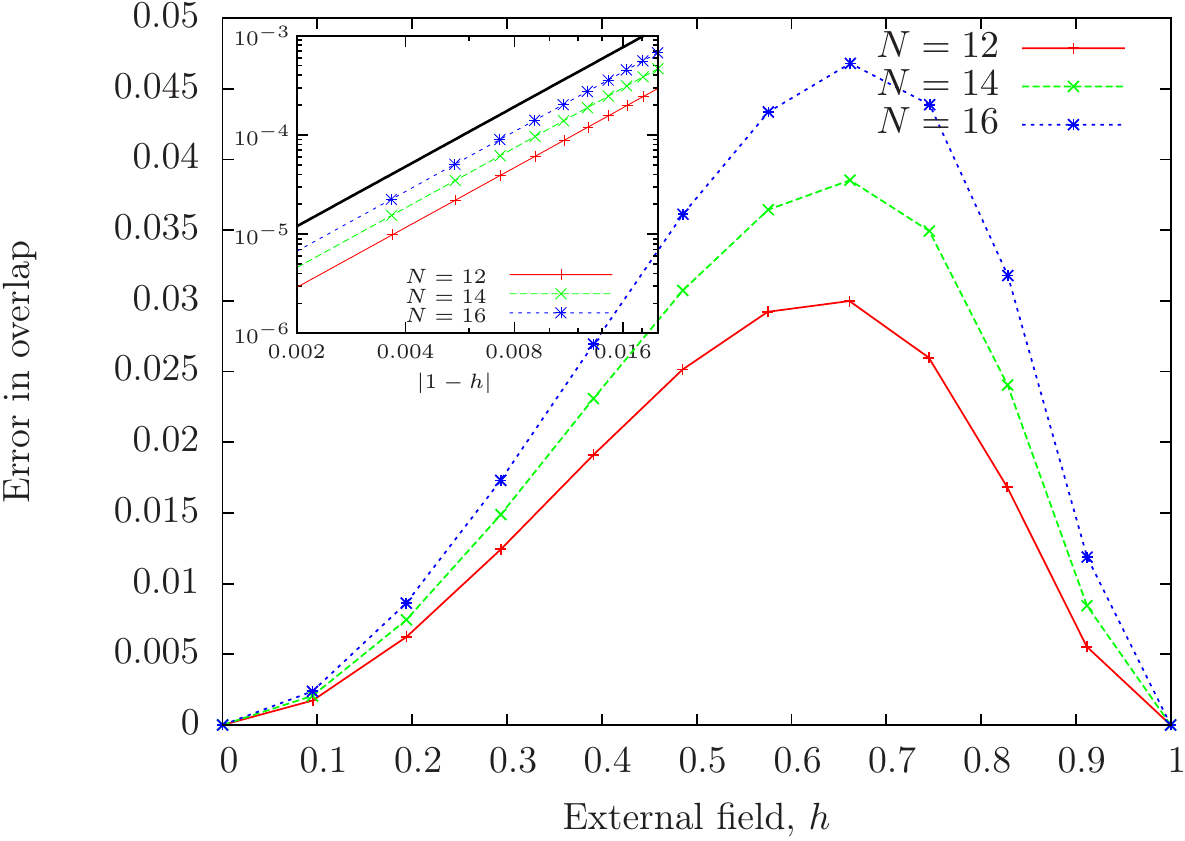}
}
    \caption{Errors of (a) the variational energy and (b) the overlap. Both plots use the value of $\delta$ that optimizes the overlap with the ground state of $H(h)$. The insets correspond to the scaling near $h_c=1$. In both cases, the error scales as $|1-h|^2$ (the black thick line serves as reference for a quadratic scaling).}
    \label{errors}
\end{figure}

First, we note that the optimal value of $h$ behaves in an almost linear fashion as a function of $\delta$, independent of the system size (Fig. \eqref{map}). However, there is a qualitative difference in the computed error for the overlap $1-\braket{gs(h)|\Psi(\delta)}$ and the expectation value of the Hamiltonian
\begin{equation}
\text{Error} = \frac{|E_\text{exact}(h)-\braket{\Psi(\delta)|H(h)|\Psi(\delta)}|}{|E_\text{exact}(h)|}.
\end{equation}
Within numerical machine precision, the optimized variational wavefunction \eqref{varWF} corresponds to the exact ground state only for $h=0$ and $h_c=1$, while errors are considerable for other values of $h$. (Fig.\eqref{errors}). (Note that $h\to\infty$ will also be exact due to the KW duality and the previous discussion.) We also see that the scaling of the errors close to $h_c=1$ is quadratic. This implies that the variational minimum is located smoothly at $\delta=0$.

%


\section{The finite critical Ising spin chain}

For $\delta = 0$, we computed the overlap of the variational wavefunction obtained from the CB \eqref{varWF} and the ground state of the (even parity) critical ITF Hamiltonian \eqref{ITFh1} for sizes up to $N=18$. In all cases, we obtain that both states are the same within machine precision. This is a remarkable result given that the expression for the CBs \eqref{CBsigmas} was obtained from the infrarred fixed point of the critical theory. It is non-trivial that it would agree with the ground state of a finite-size lattice system.

We have only been able to find an analytical proof of this statement for small sizes (see Appendix B for $N=2,3$). However, we expect this equality to hold for arbitrary sizes. Even though we cannot present a full proof, we can provide some analytical evidence that the relation holds.

Hamiltonian \eqref{ITFh1} can be solved exactly using a Jordan-Wigner (JW) transformation, followed by a Fourier transform and a Bogoliubov transformation \cite{Mussardo}. The normalized ground state can be written as
\begin{equation}
\ket{gs} = \prod_{k>0} \left(u_k+iv_kc_k^\dg c_{-k}^\dg\right)\ket{0}_c,
\label{exact gs}
\end{equation}
where $c_k^\dg$ are the fermionic creation operators,
\begin{align}
u_k =\sqrt{\frac{1+\sin(k/2)}{2}},\qquad v_k = -\sqrt{\frac{1-\sin(k/2)}{2}}, 
\end{align}
and we take $k>0$ as momenta given by (we assume $N$ to be even for simplicity)
\begin{equation}
k = \frac{\pi}{N}(2m-1), \qquad m = 1,\cdots, \frac{N}{2}.
\end{equation}

It is evident that both \eqref{varWF} and \eqref{exact gs} have very different forms. (Of course, we have to JW transform (\ref{exact gs})
back to the spin basis.) The amplitudes of the wavefunction obtained from the conformal blocks are obtained from the sum of $2^{N-1}$ terms containing square roots. On the other hand, the known exact solution is a fermionic BCS wavefunction generated from a single pairing wavefunction $g_k = v_k/u_k$ defined on half of the Brillouin zone, i.e., by $N/2$ parameters. 

In Appendix C, we prove that 
\begin{equation}
\Psi_0(\delta=0) =\braket{0|gs},
\end{equation}
for arbitrary (even) $N$. We expect that the full solution can be obtained using a similar calculation, but further work is needed to relate the combinatorics of both wavefunctions. An  alternative proof would  involve the existence of null vectors in the Virasoro modules. 
This line of research will be closer in spirit to the corresponding result for the Haldane-Shastry model, as noted above.



\section{Conclusions}

We have introduced a general framework for the construction of many-body wavefunctions for lattice systems starting from the CBs of a RCFT. Whenever present, internal symmetries can be exploited to describe the physical degrees of freedom. This was illustrated using the $SU(2)_1$ WZW model in relation to the HS spin chain \cite{iMPS, iMPS2}. In the absence of symmetry, the different fusion channels of primary fields become the natural resource for the construction of the Hilbert space. We used this idea to obtain a family of variational wavefunctions from the chiral Ising CFT. These states inherit some algebraic properties from the CFT, such as the well-known KW duality and a representation of the TLJ algebra. We provided numerical and analytical evidence that these variational wavefunctions can describe the exact ground states of the ITF spin chain for certain values of the external field, most notably the ground state at the critical point.

Further work is needed to characterize the physics of the wavefunctions obtained from the Ising CBs. In particular, it is desired to prove analytically that the homogeneous CBs $\F^{(2N)}_\pp(\delta=0)$ indeed provide the exact ground state of the critical ITF spin chain for all system sizes. A physical interpretation of this result is that the Ising degrees of freedom can be seen as the fusion channels of pairs of more elementary $\sigma$ fields, the KW duality being just the relation between two different types of pairing along the chain. Within this context, CBs that contain both $\sigma$ and Majorana fields are natural candidates for excited states \cite{workInProgress}.

In addition, it is worth studying natural generalizations to other RCFTs, such as the Potts model \cite{QuantumGroupsCFT,diFrancesco,Mussardo}, as well as lattice systems in higher dimensions. In particular, 2D systems can be studied by distributing the coordinates on the complex plane. It is known that the wavefunction obtained from $SU(2)_1$ using this configuration produces the KL state \cite{KL}. We expect then different interesting topological states from the minimal RCFTs.


\section{Acknowledgments}

We would like to thank F. Alcaraz, E. Ardonne, P. Calabrese, O. Castro-Alvaredo, 
 P. Fendley, J.I. Cirac, B. Doyon, A. Ludwig, M. Rajabpour, G. Mussardo, E. Tonni and G. Vidal 
 for useful discussions. This work was funded by grants FIS- 2012-33642, FIS-2012-38866-C05-1, and FIS2015-69167-C2-1-P 
 from the Spanish government, QUITEMAD+ S2013/ICE-2801 from the Madrid regional government and SEV-2012-0249 of the “Centro de Excelencia Severo Ochoa” Programme. SM is supported by the FPI-Severo Ochoa Ph.D. fellowship No. SVP-2013-067869. HHT acknowledges funding from the EU project SIQS.


\section{Appendix A}

Assuming that $\Psi_\pp$ is real up to a phase that does not depend on $\pp$, we have
\begin{equation}
\sum_\pp \left|\Psi_\pp\right|^2= \frac{1}{\tilde N_0^2}\sum_\pp \sum_\qq \ep_{\pp\qq}\,A_{\pp} =  \frac{2^{N-1}}{\tilde N_0^2}A_0,
\end{equation}
so that the normalization condition becomes
\begin{equation}
\tilde N_0^2 = 2^{N-1}A_0
\end{equation}
where we used
\begin{equation}
\sum_\pp \ep_{\pp\qq}= \sum_\qq \ep_{\pp\qq} = 2^{N-1}\delta_{\qq,0}.
\label{simpleIdentity}
\end{equation}
This is really convenient because the macrogroups associated to $\qq=0$ have the even numbers on one side and the odd numbers on the other. This implies that $A_0$ is just a product of sines and does not contain square roots, so that $N_0$ does not depend on $\delta$
\begin{align}
\tilde N_0^2 &= 2^{N-1}\prod_{n>m}^N \sin\left(\frac{\pi}{2N}\left(n-m\right)\right)\nonumber\\
&=2^{N-1}\left(\frac{N}{2^{N-1}}\right)^{N/2} = \frac{N^{N/2}}{2^{(N-1)(N-2)/2}}.
\end{align}
%


\section{Appendix B}

For $N=2$, it is easy to check that the exact ground state $\ket{gs}$ of \eqref{ITFh1} is given by
\begin{align}
\braket{0|gs} &= \sqrt{\frac{1+\frac{1}{\sqrt{2}}}{2}},\\
\braket{11|gs} &= \sqrt{\frac{1-\frac{1}{\sqrt{2}}}{2}}.\nonumber
\end{align}
Using the conformal block wavefunction, we have (here, $p=0,1$)
\begin{equation}
\Psi_p = \frac{1}{\sqrt{2}}\left(A_0^{(2)} + (-1)^p A_1^{(2)}\right)^{1/2}.
\end{equation}
Given that $A_0^{(2)}=1$, we compute $A_1^{(2)}$ from the macrogroup $\ell_1=(1,4)$, $\ell'_1=(2,3)$
\begin{align}
A_1^{(2)}&=\sqrt{\sin\left(\frac{\pi}{4}(4-1)\right)\sin\left(\frac{\pi}{4}(3-2)\right)}\nonumber\\
&=\sqrt{\sin\left(\frac{3\pi}{4}\right)\sin\left(\frac{\pi}{4}\right)}=\sin\left(\frac{\pi}{4}\right)=\frac{1}{\sqrt{2}}.
\end{align}
Plugging this back into the previous equation, we see that the two wavefunctions are the same.

For $N=3$, we can also compute the exact solution. We have
\begin{align}
\braket{0|gs} &=\frac{\sqrt{3}}{2},\\
\braket{110|gs}&=\braket{101|gs}=\braket{011|gs}= \frac{1}{2\sqrt{3}}.\nonumber
\end{align}
We note that all the macrogroups except for $\qq=0$ are composed of sets of one even number and two odd numbers, or viceversa. And can easily convince onceself then that
\begin{align}
\frac{A_{01}^{(3)}}{A_0^{(3)}}=\frac{A_{10}^{(3)}}{A_0^{(3)}}=\frac{A_{11}^{(3)}}{A_0^{(3)}}=\frac{\sin\left(\frac{\pi}{6}\right)}{\sin^2\left(\frac{\pi}{3}\right)}=\frac{2}{3}.
\end{align}
We can massage the CB wavefunction and write
\begin{equation}
\Psi_\pp=\frac{1}{2}\left(\frac{1}{3}+\frac{2}{3}\sum_{\qq=0}^3 \ep_{\pp\qq}\right)^{1/2}=\frac{1}{2}\left(\frac{1}{3}+\frac{8}{3}\delta_{0\pp}\right)^{1/2},
\end{equation}
which agrees with the exact ground state.


\section{Appendix C}

From the exact solution of the transverse field Ising spin chain \eqref{exact gs}, we have
\begin{align}
 \braket{0|gs}^2 =\prod_{k>0}u_k^2&= \prod_{m=1}^{N/2}\frac{1+\sin\left(\frac{\pi}{2N}(2m-1)\right)}{2}\nonumber\\
 &=\prod_{m=1}^{N/2}\frac{1+\cos\left(\frac{\pi}{2N}(2m-1)\right)}{2},
\end{align}
where $N$ is the total number of spins. This is assuming that $N$ is even.

Now, note that we can rewrite $A_\qq$ using the representation of the macrogroups 
given by the auxiliary spins \eqref{Spin macrogroup}. First, we can eliminate the square root for $\delta=0$ via the relation
\begin{align}
\prod_{j>i}^N&\sin\left[\frac{\pi}{N}\left(j-i+\frac{\alpha}{4}(s_j-s_i)\right)\right]\nonumber\\
& =\prod_{j>i}^N \sin\left[\frac{\pi}{N}\left(j-i-\frac{\alpha}{4}(s_j-s_i)\right)\right],
\label{ss} 
\end{align}
which holds for arbitrary $\alpha$ (see Appendix D). We end up with
\begin{equation}
A(\{s_k\}) = \prod_{j>i}^N \sin\left[\frac{\pi}{N}\left(j-i+\frac{1}{4}(s_j-s_i)\right)\right] . 
\label{aa} 
\end{equation}
Using the known identity for Vandermonde matrices
\begin{equation}
\sum_{\sigma\in S_N}\text{sgn}(\sigma)\prod_{n=1}^N \alpha_n^{\sigma(n)-1} = \prod_{n>m}^N (\alpha_n-\alpha_m)
\end{equation}
and equation \eqref{ComplexToSine}, this can be written as
\begin{align}
A(\{s_k\})=C_N\sum_{\sigma\in S_N}\text{sgn}(\sigma)\left(\prod_{j=1}^N a_{j,\sigma(j)}\right)\left(\prod_{j=1}^N b_{j,\sigma(j)}\right)
\end{align}
where $C_N = (2i)^{-N(N-1)/2}e^{-i\frac{\pi}{2}(N^2-1)}$ is a global constant,
\begin{equation}
a_{r,t} = \exp\left(i\frac{2\pi}{N}r(t-1)\right)
\end{equation}
%
and
\begin{equation}
b_{r,t} = \exp\left(i\frac{\pi}{4N}s_r(2t-N-1)\right)
\end{equation}
contains all the dependence on $\{s_k\}$. We can perform the sum
\begin{align}
\sum_{\{s_k\}}\prod_{j=1}^N b_{j,\sigma(j)}&=\prod_{j=1}^N\left(2\cos\left[\frac{\pi}{4N}\left(2\sigma(j)-N-1\right)\right]\right)\nonumber\\
&=2^N\prod_{j=1}^N\cos\left[\frac{\pi}{4N}\left(2j-N-1\right)\right].
\end{align}
In other words, the auxiliary spins decouple from the permutation. We can recombine the other part of the equation using once again the Vandermonde identity
\begin{align}
C_N&\left[\sum_{\sigma\in S_N}\text{sgn}(\sigma)\left(\prod_{j=1}^N  a_{j,\sigma(j)}\right)\right]\nonumber\\
&=\prod_{j>i}^N \sin\left[\frac{\pi}{N}\left(j-i\right)\right]= \left(\frac{N}{2^{N-1}}\right)^{N/2},
\end{align}
so that
\begin{align}
\sum_{\{s_k\}}&A(\{s_k\})= N_0^2\prod_{j=1}^N\cos\left[\frac{\pi}{4N}\left(2j-N-1\right)\right].
\end{align}
We have then
\begin{equation}
\Psi_0^2(\delta=0) = \prod_{m=1}^N\cos\left[\frac{\pi}{4N}\left(2m-N-1\right)\right].
\end{equation}
Using the relation
\begin{equation}
\prod_{m=1}^N\cos\left[\frac{\pi}{4N}\left(2m-N-1\right)\right] = \prod_{m=1}^{N/2}\cos^2\left[\frac{\pi}{4N}\left(2m-1\right)\right],
\end{equation}
and the identity
\begin{equation}
\cos^2(\theta) \equiv \frac{1+\cos(2\theta)}{2},
\end{equation}
we get
\begin{align}
\prod_{m=1}^N & \cos\left[\frac{\pi}{4N}\left(2m-N-1\right)\right] \\
&= \prod_{m=1}^{N/2}\frac{1+\cos\left(\frac{\pi}{2N}(2m-1)\right)}{2},\nonumber
\end{align}
so that
\begin{equation}
\Psi_0^2(\delta=0) =\braket{0|gs}^2.
\end{equation}
%


\section{Appendix D}

Below we prove  the relation (\ref{ss}). 
Consider the complex function
\begin{equation}
f(z) = \prod_{j>i}^N\sin\left[\frac{\pi}{N}\left(j-i\right)+\frac{z}{2}(s_j-s_i)\right],
\end{equation}
where $s_i=\pm 1$, for all $i=1,\cdots,N$. Given that $s_j - s_i =0, \pm 2$, note that $f(z+2\pi) = f(z)$. Also, being the product of analytic functions, $f(z)$ is also analytic on the whole complex plane.

We would like to prove that $f(z)$ is an even function. This holds trivially if $s_j=1$ for all $j=1,\cdots,N$. For the general case, let us define the sets
\begin{equation}
A_\pm = \{j\, |\, s_j=\pm 1\},
\end{equation}
so that $A_+\cup A_- = \{1,\cdots, N\}$. Using this notation, we can write the ratio
\begin{align}
\frac{f(z)}{f(-z)} = \prod_{j\in A_+}\left[\prod_{\substack{i\in A_- \\ i<j}} \frac{\sin\left(\frac{\pi}{N}(N+i-j)-z\right)}{\sin\left(\frac{\pi}{N}(j-i)-z\right)}\right.\nonumber\\
\left.\prod_{\substack{i\in A_- \\ i>j}} \frac{\sin\left(\frac{\pi}{N}(i-j)-z\right)}{\sin\left(\frac{\pi}{N}(N+j-i)-z\right)}\right].
\label{ratioff}
\end{align}
Let us now define the (non-symmetric) functions
\begin{align}
d_R(i,j) &=  \left\{
  \begin{array}{l l}
    i-j & \quad i\geq j,\\
    N+i-j & \quad i < j,
  \end{array} \right. \\
d_L(i,j) &=  \left\{
  \begin{array}{l l}
    N+j-i & \quad i > j,\\
    j-i & \quad i \leq j.
  \end{array} \right.
\end{align}
These functions can be interpreted geometrically. Assume the integers $\{1,\cdots, N\}$ are evenly distributed on a circle in a clockwise ascending order. Then, $d_R(i,j)$ (respectively $d_L(i,j)$) is the distance from $j$ to $i$ going only in the clockwise (respectively, anticlockwise) direction. (Note that $d_R(i,j) = d_L(j,i)$.)

We can then write \eqref{ratioff} as
\begin{equation}
\frac{f(z)}{f(-z)} = \prod_{j\in A_+}\prod_{i\in A_-} \frac{\sin\left(\frac{\pi}{N}d_R(i,j)-z\right)}{\sin\left(\frac{\pi}{N}d_L(i,j)-z\right)}.
\end{equation}
In this formulation, $f(z)$ will be an even function if the lists of integers
\begin{align}
R &= (d_R(i,j) \,| \, i\in A_-, j\in A_+), \nonumber\\
L &= (d_L(i,j) \,| \, i\in A_-, j\in A_+)
\label{DefRL}
\end{align}
contain the same elements with the same multiplicities. In other words, we must prove that the set of all the distances from every element of $A_+$ to every element in $A_-$ counting clockwise is the same as the list counting anticlockwise.

In order to prove this statement, note that if $d_R(i,j)=r$, then $d_L(i,j) = N-r$. This implies that $R$ and $L$ will be equal if we can pair the elements within the same list as $(r,N-r)$. (If $N$ is even, this statement is true except if $r=N/2$. In that case, the element is trivially in both lists and does not need pairing.) We will then focus on pairing the elements in list $R$.

Consider the matrix defined by
\begin{equation}
[D(P,Q)]_{i,j} = d_R(p_i,q_j),
\end{equation}
where $P=\{p_i\}$ and $Q=\{q_i\}$ are two subsets of $\{1,\cdots,N\}$. We will say $D(P,Q)$ is a \emph{balanced} matrix if there are the same number of matrix elements that take the value $r$ (with $r\neq 0$) and $N-r$. (Once again, if $N$ is even, we also need $r\neq N/2$.)

It is easy to see that $D(A_+,A_+\cup A_-)$ is a balanced matrix because we can always pair $d_R(i,j)$ with $d_R(i,N-j)$. Likewise, $D(A_+,A_+)$ is also balanced because $d_R(i,j)$ pairs with $d_R(j,i)$. We have then that $D(A_+,A_-)$ is a submatrix of $D(A_+,A_+\cup A_-)$ than can be obtained by removing a balanced submatrix. This implies that $D(A_+,A_-)$ is balanced.

Using this result, we see that for every $i\in A_+$, $j\in A_-$ such that $d_R(i,j)=r$, there exist $i'\in A_+$, $j'\in A_-$ such that $d_R(i',j') = N-r$, or equivalently, $d_L(i',j')=r$. So both $R$ and $L$ are equal and
\begin{equation}
f(z) = f(-z),
\end{equation}
which is what we wanted to prove. 

\end{document}